\documentclass[11pt]{article}

\newcommand{\be}{\begin{equation}}
\newcommand{\ee}{\end{equation}}
\def\onehalf{\textstyle{\frac{1}{2}}}
\def\D{{\mathcal D}{}}

\def\Abol{{\stackrel{~\circ}{A}}{}}

\begin{document}
\renewcommand{\thefootnote}{\fnsymbol{footnote}}
\noindent
{\Large \bf Torsion as Alternative to Curvature in the}
\vskip 0.3cm \noindent
{\Large \bf Description of Gravitation}\footnote{Talk
presented at {\it Fourth International Winter Conference on Mathematical Methods 
in Physics}, 09 - 13 August 2004, Centro Brasileiro de Pesquisas F\'{\i}sicas 
(CBPF/MCT), Rio de Janeiro, Brazil.}
\vskip 0.7cm
\noindent
{\bf V. C. de Andrade}\footnote{Speaker.} \hfill \\
{\it Universidade de Bras\'\i lia \hfill \\
Bras\'\i lia DF, Brazil}
\vskip 0.2cm \noindent
{\bf H. I. Arcos\footnote{Permanent address: Universidad Tecnol\'ogica de Pereira, 
A.A. 97, La
Julita, Pereira, Colombia.} and J. G. Pereira} \hfill \\
{\it Instituto de F\'{\i}sica Te\'orica, UNESP \hfill \\
S\~ao Paulo SP, Brazil}

\vskip 1.0cm

\begin{abstract}
\noindent
The general covariance principle, seen as an {\em  active} version of the  principle of
equivalence, is used to study the gravitational coupling prescription in the presence of 
curvature and torsion. It is concluded that the coupling prescription determined by this
principle is  always equivalent with the corresponding prescription of general relativity.
An application to the  case of a Dirac spinor is made.
\end{abstract}

\section{Introduction}
\setcounter{footnote}{0}
\renewcommand{\thefootnote}{\arabic{footnote}}

According to the teleparallel equivalent of general relativity,  curvature and torsion are
alternative ways of describing the gravitational field,  and consequently related to the
same degrees of freedom of gravity. However, more  general gravity theories \cite{blago},
like for example Einstein-Cartan and gauge  theories for  the Poincar\'e and the affine
groups \cite{gauge}, consider curvature and torsion  as representing independent degrees of
freedom. In these models, differently from  teleparallel gravity, torsion becomes relevant
only when spins are important 
\cite{gh96}. According to this point of view, torsion represents additional  degrees of
freedom in relation to curvature, and consequently new physics  phenomena might be
associated with it \cite{shapiro}.

The above described difference rises a conceptual question on the actual role  played by
torsion in the description of the gravitational interaction. This  question shows up due to
the difficulty in determining the correct form of the  gravitational coupling
pre\-scrip\-tion in the presence of curvature and torsion.  In fact, differently from all
other interactions of nature, where the requirement of covariance does determine the gauge
connection, in the presence of  curvature and torsion, covariance---seen as a consequence
of the strong  equivalence principle \cite{weinberg}---is not able to determine the form of
the  gravitational coupling prescription. The reason for this indefiniteness is that  the
space of Lorentz connections is an affine space \cite{KoNo}, and consequently  one can
always add a tensor (actually a Lorentz-valued covector) to a given  connection without
destroying the covariance of the theory. Notice that in the  specific cases of general
relativity and teleparallel gravity, characterized  respectively by a vanishing torsion and
a vanishing curvature, the above  indefiniteness is absent since in these cases the
connections are uniquely  determined---and the corresponding coupling prescriptions
completely specified--- by the combined use of covariance and the strong equivalence
principle. Notice  furthermore that in the case of internal (Yang-Mills) gauge theories,
where the concept of torsion  is absent\footnote{ We remark that absence of torsion, like
in internal gauge theories, is different from the presence of a vanishing torsion, which
happens in  general relativity \cite{livro}.}, the above indefiniteness is not present
either.

A possible answer to this problem can be obtained by using the so called {\em  principle of
general covariance}, an active version of the strong equivalence  principle \cite{ap04}.
According to this principle, in order to make an equation  generally covariant, a
connection is always necessary, which is in principle  concerned only with the {\em
inertial} properties of the coordinate system under consideration. Then, by using the
equivalence between inertial and gravitational  effects, instead of representing inertial
properties, this connection can  equivalently be assumed to represent a {\em true
gravitational field}. In this  way, equations valid in the presence of gravitation are
obtained from the  corresponding special relativity equations. It is important to remark
that general  covariance by itself is empty of physical content as any equation can be {\it 
made} generally covariant. Only when use is made of the strong equivalence principle,  and
the inertial compensating term is interpreted as representing a true  gravitational field,
the principle of general covariance can be seen as an  alternative version of the strong
equivalence principle. Now, when the {\em purely  inertial} connection is replaced by a
connection representing a {\em true  gravitational field}, the principle of general
covariance naturally defines a  covariant derivative, and consequently also a gravitational
coupling prescription.  The purpose of the present work will be to use this principle to
determine the  form of the gravitational coupling prescription in the presence of both
curvature  and torsion.

\section{General covariance principle and coupling prescription}

The process of obtaining a gravitational coupling prescription from the general  covariance
principle comprises two steps. The first is to pass to a general  nonholonomic frame, where
inertial effects---which appear in the form of a  connection, or compensating term---are
present. Then, by using the strong  equivalence principle, instead of inertial effects, the
compensating term can be  replaced by a connection representing a true gravitational field.
In this way, a  covariant derivative, and consequently a gravitational coupling
prescription,  is  obtained. Let us then see how the whole process works.

\subsection{General frames}

Let us consider the Minkowski spacetime\footnote{We use the Greek alphabet $\mu, \nu, \rho,
\dots = 0, 1, 2, 3$ to denote spacetime indices, and the Latin alphabet $a, b, c, \dots =
0, 1, 2, 3$ to denote indices related to the (local) tangent Minkowski spaces.} of special
relativity, endowed with the Lorentzian metric $\eta$. In this spacetime one can take the
frame $\delta_a=\delta_a{}^\mu\partial_\mu$ as being a trivial (holonomous) tetrad, with
components $\delta_a{}^\mu$. Consider now a {\em local}, that is, point-dependent  Lorentz
transformation $\Lambda_a{}^b=\Lambda_a{}^b(x)$. It yields the new frame $h_{a} = 
h_{a}{}^{\mu} \partial_\mu$, with components $h_a{}^\mu \equiv h_a{}^\mu(x)$ given by
\begin{equation}
h_{a}{}^{\mu} = \Lambda_{a}{}^{b} \; \delta_{b}{}^{\mu }.
\label{deltabar}
\end{equation}
Notice that, on account of the locality of the Lorentz transformation,
the new frame $h_a$ is nonholonomous, $[h_b, h_c] = f^a{}_{bc} \, h_a$, with the 
coefficient of nonholonomy given by
\begin{equation}
f^a{}_{bc} = h_b{}^\mu \, h_c{}^\nu (\partial_\nu
h^a{}_\mu - \partial_\mu h^a{}_\nu).
\end{equation}
Making use of the orthogonality property of the tetrads, therefore, we see from
Eq.~(\ref{deltabar}) that the Lorentz group element can be written in the  form
$\Lambda_b{}^d = h_b{}^\rho \delta_\rho{}^d$. From this expression, it follows  that
\begin{equation}
\Lambda^c{}_d \, (h_a \Lambda_b{}^d) =
\textstyle{\frac{1}{2}} \left(f_b{}^c{}_a + f_a{}^c{}_b - f^c{}_{ba} \right).
\label{deltabar e f}
\end{equation}

Let us consider now a vector field $v^c$ in the Minkowski spacetime. Its ordinary 
derivative in the frame $\delta_a$ is
\begin{equation}
\partial_a v^c = \delta_a{}^\mu \partial_\mu v^c.
\label{hder}
\end{equation}
Under a local Lorentz transformation, the vector field transforms according to 
$V^d = \Lambda^d{}_c \, v^c$, and it is easy to see that $\partial_a v^c = 
\Lambda^b{}_a \, \Lambda_d{}^c {\mathcal D}_b V^d$, where
\begin{equation}
{\mathcal D}_a V^c = h_a V^c +
\Lambda^c{}_d \, (h_a \Lambda_b{}^d) \; V^b.
\label{hder2}
\end{equation}
In the frame $h_a$, therefore, using the identity (\ref{deltabar e f}), the 
derivative (\ref{hder2}) acquires the form
\begin{equation}
{\mathcal D}_a V^c = h_a V^c + \onehalf \left(f_b{}^c{}_a + f_a{}^c{}_b - 
f^c{}_{ba} \right) V^b.
\label{nhder3}
\end{equation}
The freedom to choose any tetrad $\{h_{a}\}$ as a moving frame in the Minkowski  spacetime
introduces the compensating term $\frac{1}{2} (f_b{}^c{}_a + f_a{}^c{}_b  - f^c{}_{ba} )$
in the derivative of the vector field. This term, of course, is  concerned only with the
inertial properties of the frame.

\subsection{Equivalence between inertia and gravitation: gravitational coupling 
prescription}

According to the general covariance principle, the derivative valid in the presence of
gravitation can be obtained from the corresponding Minkowski covariant  derivative by
replacing the inertial compensating term by a connection $A^c{}_{ab}$ representing a true
gravitational field. Considering a general  Lorentz-valued connection presenting both
curvature and torsion, one can always write~\cite{abp1}
\begin{equation}
A^c{}_{ba} - A^c{}_{ab} = T^c{}_{ab} + f^c{}_{ab},
\end{equation}
with $T^c{}_{ba}$ the torsion of the connection $A^c{}_{ab}$. Use of this
equation for three different combination of indices gives
\begin{equation}
\onehalf \left( f_b{}^c{}_a + f_a{}^c{}_b - f^c{}_{ba} \right) =
A^c{}_{ab} - K^c{}_{ab}.
\label{equiva}
\end{equation}
where
\begin{equation}
K^c{}_{ab} = \onehalf \left( T_b{}^c{}_a + T_a{}^c{}_b - T^c{}_{ba} \right)
\label{contor}
\end{equation}
is the contortion tensor. Equation (\ref{equiva}) is completely general, and is  the
crucial point of the approach. It is actually an expression of the equivalence principle in
the sense that, whereas its left-hand side involves only  {\it inertial} properties of the
frames, its right-hand side contains purely {\it gravitational} quantities. Using this
expression, the derivative (\ref{nhder3})  becomes
\begin{equation}
{\mathcal D}_a V^c = h_a V^c + \left(A^c{}_{ab} - K^c{}_{ab} \right) V^b \equiv
h_a{}^\mu {\mathcal D}_\mu V^c,
\label{nhder4}
\end{equation}
where
\begin{equation}
\D_\mu V^c =
\partial_\mu V^c + (A^c{}_{a\mu} - K^c{}_{a\mu}) \; V^a
\label{gfic}
\end{equation}
is a generalized Fock--Ivanenko derivative. Using then the vector representation 
$(S_{ab})^c{}_d $ of the Lorentz generators~\cite{ramond}, the generalized Fock--
Ivanenko derivative (\ref{gfic}) can be written in the form
\begin{equation}
\D_{\mu} X^{c} =
\partial_{\mu} X^{c} - \textstyle{\frac{i}{2}} (A^{ab}{}_{\mu} - K^{ab}{}_{\mu}) 
\;
(S_{ab})^c{}_d \; X^d.
\label{acomivetor2}
\end{equation}

Now, although obtained in the case of a Lorentz vector field, the compensating term
(\ref{deltabar e f}) can be easily verified to be the same for  any field. In fact,
denoting by $U \equiv U(\Lambda)$ the element of the Lorentz group  in an arbitrary
representation, it can be shown that~\cite{mospe}
\begin{equation}
(h_{a} U) U^{-1} = - \textstyle{\frac{i}{4}}
\left(f_{bca} + f_{acb} - f_{cba} \right) \, J^{bc},
\label{vacua}
\end{equation}
with $J^{bc}$ denoting the corresponding Lorentz generator. In the case of fields 
carrying an arbitrary representation of the Lorentz group, therefore, the 
covariant
derivative (\ref{acomivetor2}) acquires the form
\begin{equation}
\D_{\mu} = \partial_{\mu} - \textstyle{\frac{i}{2}} \left(A^{ab}{}_{\mu} - 
K^{ab}{}_{\mu} \right) J_{ab}.
\label{genecova1}
\end{equation}
This means that, in the presence of curvature and torsion, the gravitational  coupling
prescription implied by the general covariance principle amounts to replace
\begin{equation}
\partial_a \equiv \delta^\mu{}_a \partial_\mu \rightarrow
\D_a \equiv h^\mu{}_a \D_\mu.
\label{genecova2}
\end{equation}
We notice finally that, due to the relation
\begin{equation}
A^{ab}{}_{\mu} - K^{ab}{}_{\mu} = \Abol^{ab}{}_{\mu},
\end{equation}
with $\Abol^{ab}{}_{\mu}$ the spin connection of general relativity, the above 
coupling prescription is clearly equivalent with the coupling prescription of 
general relativity.

\section{Example: the spinor field}

The gravitational coupling prescription (\ref{genecova1}-\ref{genecova2}) has  already been
applied to study the motion of both a spinless and a spinning  particle~\cite{newview}.
Here, we apply it to the case of a Dirac spinor in the  presence of curvature and torsion.

\subsection{Dirac equation}

The Dirac equation in flat spacetime can be obtained from the Lagrangian (we use  units in
which $\hbar = c = 1$)
\begin{equation}
\mathcal{L} = \textstyle{\frac{i}{2}} \Big(
\bar{\psi}\gamma^a\delta_a{}^\mu \partial_\mu \psi -
\partial_\mu \bar{\psi}\gamma^a\delta_a{}^\mu \psi
\Big) - m \, \bar{\psi}\psi,
\label{lagrDirac0}
\end{equation}
where $\delta_a{}^\mu$ is a trivial tetrad, $m$ is the particle's mass, and 
$\{\gamma^{a}\}$ are (constant) Dirac matrices in a given representation. Making 
use of the coupling prescription  (\ref{genecova1}-\ref{genecova2}), with $J^{bc} 
= {\sigma^{bc}}/{2}:= ({i}/{4}) [\gamma^{b},\gamma^{c}]$ the spinor representation 
of the Lorentz generators, we obtain
\begin{equation}
\mathcal{L} = \textstyle{\frac{i}{2}} \Big(
\bar{\psi} h_a{}^\mu\gamma^a \D_\mu \psi -
\D_\mu \bar{\psi} h_a{}^\mu\gamma^a \psi
\Big) - m \, \bar{\psi}\psi,
\label{diraclagr}
\end{equation}
where the Fock-Ivanenko derivative operator is given by
\begin{equation}
\D_{\mu}\psi =
\partial_{\mu}\psi - \textstyle{\frac{i}{4}} \; (A^{bc}{}_{\mu} - K^{bc}{}_{\mu} ) 
\,{\sigma_{bc}} \psi.
\label{fi}
\end{equation}
This covariant derivative yields the coupling prescription for spin-1/2 fields in 
the presence of curvature and torsion. As usual, a functional derivative with 
respect to $A^{bc}{}_{\mu} - K^{bc}{}_{\mu} \equiv \Abol^{bc}{}_{\mu}$ yields the 
spin tensor. A straightforward calculation shows that the Dirac Lagrangian 
(\ref{diraclagr})
gives rise to
\begin{equation}
i \gamma^a h_a{}^\mu \D_\mu \psi = m \, \psi,
\label{diracequation}
\end{equation}
which is the Dirac equation in the presence of curvature and torsion. Of course,  as
already mentioned, it is equivalent with the Dirac equation in the context of general
relativity \cite{dirac}.

\subsection{Torsion decomposition}

As is well known, torsion can be decomposed in irreducible
components under the global Lorentz group \cite{hs}
 \be
 T_{\lambda
\mu \nu} = \textstyle{\frac{2}{3}} \left(t_{\lambda \mu \nu} -
t_{\lambda \nu \mu} \right) + \frac{1}{3} \left(g_{\lambda \mu}
T_\nu - g_{\lambda \nu} T_\mu \right) + \epsilon_{\lambda \mu \nu
\rho} \, S^\rho.
\label{deco}
\ee
In this expression, $T_\mu$ and $S^\rho$ represent the vector and axial parts of 
torsion, defined respectively by
\be
T_{\mu} =  T^{\nu}{}_{\nu \mu} \quad \mbox{and} \quad S^{\mu} =
\textstyle{\frac{1}{6}} \epsilon^{\mu\nu\rho\sigma} \,
T_{\nu\rho\sigma},
\label{pt3}
\ee
and $t_{\lambda \mu \nu}$ is the purely tensor part, which satisfies the 
properties $t_{\lambda \mu \nu} = t_{\mu \lambda \nu}$ and $t^\mu{}_{\mu \lambda } 
= 0 = t^\mu{}_{\lambda \mu}$. As a simple calculation shows,
\be \textstyle{\frac{i}{4}} \, K^{bc}{}_a \, \gamma^{a} \,
\sigma_{bc} = - \gamma^{a} \left( \frac{1}{2} T_{a} + \frac{3
i}{4} \, S_{a} \, \gamma^{5}  \right), \label{deco22} \ee with
$\gamma^{5}=\gamma_{5}:=i\gamma^{0}\gamma^{1}\gamma^{2}\gamma^{3}$.
The covariant derivative (\ref{fi}) then becomes
\begin{equation}
\D_{\mu}\psi =
\left( \partial_{\mu} - \textstyle{\frac{i}{4}} \; A^{bc}{}_{\mu}\,{\sigma_{bc}} - 
\frac{1}{2} \, T_\mu - \frac{3 i}{4} \, S_\mu \, \gamma^{5}  \right) \psi.
\label{fi2}
\end{equation}
We observe that, whereas the functional derivative of the Lagrangian 
(\ref{diraclagr}) in relation to the connection $A^{bc}{}_{\mu}$ still gives the 
spin tensor, derivatives with respect to $T_\mu$ and $S_\mu$ give respectively the 
vector and the axial-vector currents of the spinor field.

Substituting now the covariant derivative (\ref{fi2}) in the equation 
(\ref{diracequation}), we get
\begin{equation}
i \, \gamma^{\mu} \left(
\partial_{\mu} - \textstyle{\frac{i}{4}} \; A^{bc}{}_{\mu}\,{\sigma_{bc}} - 
\frac{1}{2} \, T_{\mu} - \frac{3 i}{4} \; S_{\mu} \,\gamma^{5}
\right)  \psi = m \psi,
\label{diracgeneral}
\end{equation}
where $\gamma^{\mu} \equiv \gamma^{\mu}(x) = \gamma^a \, h_a{}^\mu$. This is the 
Dirac equation in the presence of curvature and torsion, written in terms of 
irreducible components for torsion. In the specific case of teleparallel gravity, 
$A^{bc}{}_{\mu}=0$, and the resulting Dirac equation turns out to be written in 
terms of the vector and axial-vector torsions only \cite{mospe}. We remark that in 
the general relativity case, where the Fock-Ivanenko derivative is given by 
$\D_{\mu}\psi = \partial_{\mu}\psi - \textstyle{\frac{i}{4}} \; \Abol^{bc}{}_{\mu} 
\, {\sigma_{bc}} \psi$, if the spin connection $\Abol^{bc}{}_{\mu}$ is written in 
terms of the coefficient of nonholonomy $f^a{}_{bc}$, a decomposition similar to 
(\ref{deco22}) can be made, and the Dirac equation turns out to be written in 
terms of the trace and the pseudo-trace of $f^a{}_{bc}$ only\footnote{We thank R. 
Aldrovandi for calling our attention to this point.}.

\section{Final Comments}

A fundamental difference between general relativity and teleparallel gravity is  that,
whereas in the former curvature is used to geometrize the gravitational 
interaction---spinless particles follow geodesics---in the latter torsion describes the
gravitational interaction by acting as a force---trajectories are  not given by geodesics,
but by force equations \cite{glf}. According to the  teleparallel approach, therefore, the
role played by torsion is quite well  defined: it appears as an alternative to curvature in
the description of the  gravitational field, and is consequently related with the same
degrees of freedom  of gravity. Now, this interpretation is completely different from that
appearing in more general theories, like Einstein--Cartan and gauge theories for the 
Poincar\'e and the affine groups. In these theories, curvature and torsion are  considered
as independent fields, related with different degrees of freedom of  gravity, and
consequently with different physical phenomena. This is a conflicting  situation as these
two interpretations cannot be both correct.

As an attempt to solve the above described paradox, we have used the general  covariance
principle---seen as an alternative version of the strong equivalence  principle---to study
the gravitational coupling prescription in the presence of  curvature and torsion.
According to this principle, the dynamical spin connection,  that is, the spin connection
defining the covariant derivative, and consequently  the gravitational coupling
prescription, is $A^c{}_{ab} - K^c{}_{ab}$. Even in the presence of curvature and torsion,
therefore, torsion appears as playing the role  of gravitational force. This result gives
support to the point of view of  teleparallel gravity, according to which torsion does not
represent additional  degrees of freedom of gravity, but simply an alternative way of
representing the  gravitational field. Furthermore, since $A^c{}_{ab} - K^c{}_{ab} = 
\Abol^c{}_{ab}$, the ensuing coupling prescription will always be equivalent with  the
coupling prescription of general relativity, a result that reinforces the  completeness of
this theory.

It is important to add that, at least up to now, there are no compelling  experimental
evidences for {\it new physics associated with torsion}. We could  then say that the
teleparallel point of view is favored by the available  experimental data. For example, no
new gravitational physics has ever been  reported near a neutron star. On the other hand,
it is true that, due to the weakness of the  gravitational interaction, no experimental
data exist on the coupling of the spin  of the fundamental particles to gravitation.
Anyway, precision  experiments \cite{exp} either in laboratory or as astrophysical and
cosmological  tests are expected to be available in the foreseeable future, when then a
final  answer will hopefully be achieved.

\section*{Acknowledgments}
The authors would like to thank R. Aldrovandi and R. A. Mosna for very useful 
discussions. They would like also to thank FINATEC (UnB), COLCIENCIAS (Colombia), 
CAPES, CNPq and FAPESP for partial financial support.

\end{document}